# TRANSMITTANCE QUANTITIES PROBABILITY DISTRIBUTIONS OF WAVES THROUGH DISORDERED SYSTEMS

**Gabriel Cwilich and Fredy R Zypman**

YESHIVA UNIVERSITY
Department of Physics
New York, NY

**ABSTRACT**

Waves propagate through disordered systems in a variety of regimes. There is a threshold of disorder beyond which waves become localized and transport becomes restricted. The intensity I of the wave transmitted through a system has a dependence on the length L of the sample that is characteristic of the regime. For example, I decays as $L^{-1}$ in the diffusive regime. It is of current interest to characterize the transport regime of a wave from statistical studies of the transmittance quantities through it. Studies suggest that the probability distribution of the intensity could be used to characterize the localized regime.[1] There is an ongoing debate on what deviations from the classical Rayleigh distribution are to be expected. In this numerical work, we use scalar waves to obtain the intensity, transmission, and conductance of waves through a disordered system. We calculate the intensity, by setting an incoming plane wave towards the sample from a fixed direction. The outgoing intensity is then calculated at one point in space. This process is repeated for a collection of samples belonging to the same ensemble that characterizes the disorder, and we construct the probability distribution of the intensity. In the case of transmission, we evaluate the field arriving to a series of points distributed in the far field, and repeat the same statistical analysis. For the conductance, we calculate the field at the same series of points for incoming waves in different directions. We analyze the distribution of the transmittance quantities and their change with the degree of disorder.

## INTRODUCTION

The propagation of waves in disordered systems has interested scientists since Lord Rayleigh studied the diffusion of light in the atmosphere to explain the color of the sky[2], and led to the development of the theory of Radiative Transfer[3]. A pivotal advance was the work of Anderson[4] raising the possibility that disorder can lead to non-diffusive behavior in which the intensity transmitted decreases exponentially as a function of the length of the sample — the so-called localized regime. New theoretical ideas like the scaling theory of localization[5], weak localization[6,7], universal conductance fluctuations[8] and Wigner dwelling times[9,10], were followed, and the new field of Mesoscopic Physics reached and influenced many experimental areas: among them electronic systems[11,12], microwaves[13], optics[14,15], acoustics[16], geophysics[17], laser physics[18,19], medical physics[20] and atomic physics[21]. One particular problem that remains central to this field is to understand the signature of the propagation of a signal in the different regimes (after all localization is the absence of transmission!), since disagreement persists about the interpretation of experimental results[22]. Theoretical analyses of certain characteristics of the propagation, in particular its statistical properties[23,24], are of great interest, since those properties are now becoming experimentally accessible[25,26]. It has been argued recently that the distribution of the transmission intensity through a random structure can provide insights on the nature of the propagation inside the structure[27].

**DESCRITPION OF THE SYSTEM**

The system is completely randomized, with the degree of disorder characterized by the filling fraction of scatterers.

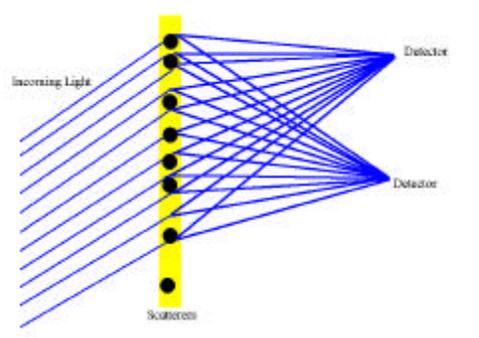

Figure 1

Figure 1 shows the parameters of a typical configuration. An incoming plane wave with direction $\vec{k}_a$ impinges on the system. A detector is placed along the direction $\vec{k}_b$, where the scattered wave is measured. The square of the magnitude of the electric field for a fixed direction is the intensity $I_{ab}$, ($I_{ab}=|E_{ab}|^2$). The transmission is defined by $T_a=|\sum_b E_{ab}|^2$, that is, the intensity collected from many detectors. Finally, the conductance is $\sigma = |\sum_{ab} E_{ab}|^2$, corresponding to the experimental situation of many sources and many detectors.

In Scalar Diffraction Theory, each point of the openings emits a secondary, Hüygens wave. The field is given by Kirkchoff's expression $E = \sum \frac{e^{ikr}}{r}(\cos\alpha + \cos\theta)$, where α is the incident angle of the plane wave, θ is the angle of the outgoing wave, k is the wave number, $r$ is the distance from a point on the surface of a scatterer to a given detector, and the sum runs over the surface of the scatterers. We will obtain numerically, for various filling fractions, the statistical distribution of all the transmittance quantities mentioned above.

**NUMERICAL RESULTS**

We produced histograms of $I_{ab}$, $T_a$, and σ, for filling factors between 20% and 90% --the filling fraction is defined as 100x(volume of single scatterer)x(number of scatterers)/(volume of container). For each histogram we considered 64,000 different configurations of the ensemble. For each configuration we calculated the desired transmittance quantity.

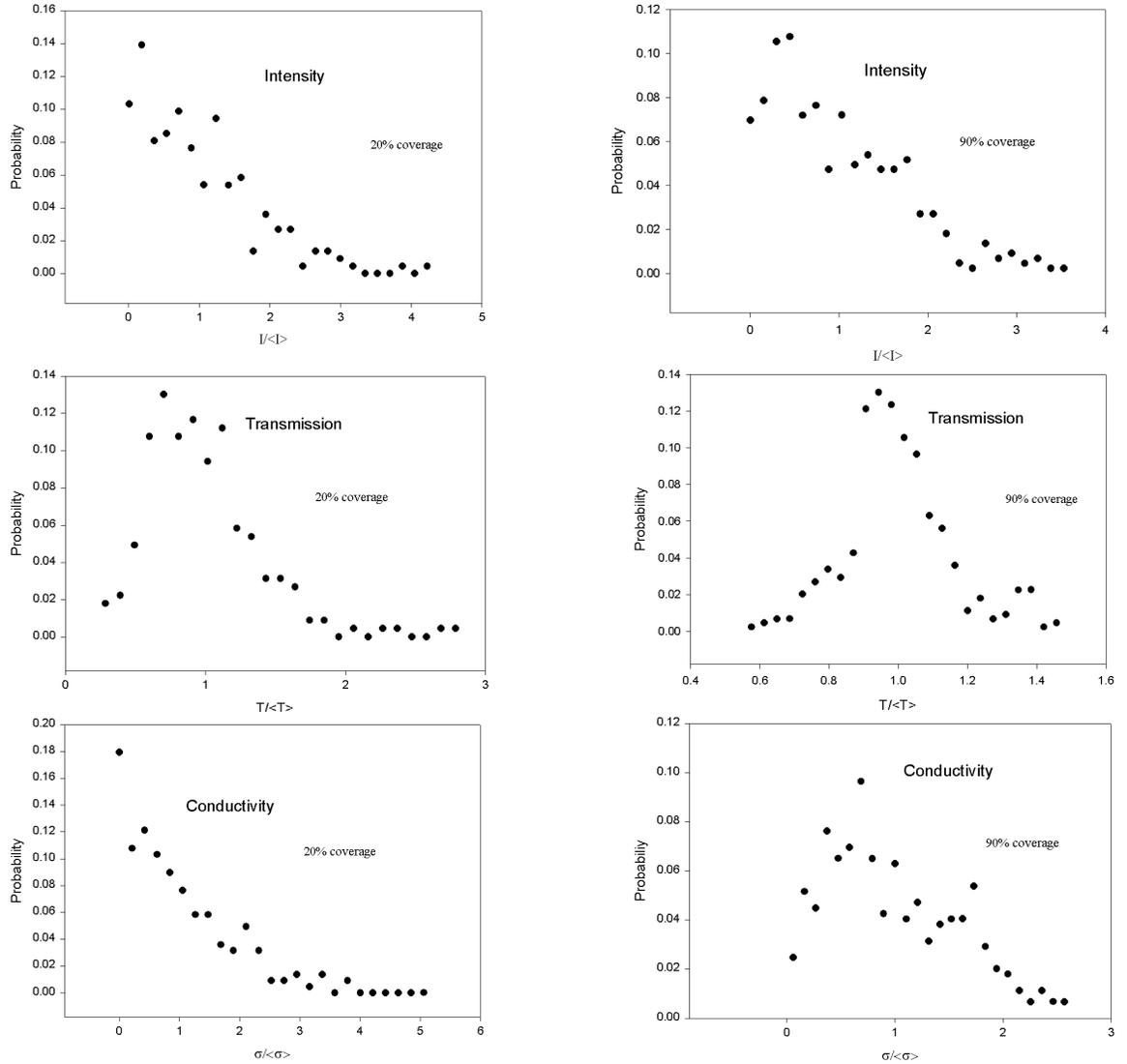

Figure 2. The left column corresponds to 20% coverage, and the right one to 90%.

In Figure 2, we present the probability distributions for all transmittance quantities for the lowest and highest filling factors considered. Qualitatively, the Intensity curves are decaying presenting the long tails characteristic of Rayleigh distributions. We will see that, quantitatively, they do not follow a Rayleigh statistics but a hybrid statistics between a ballistic and a diffusive regime (Rayleigh's corresponds to the purely diffusive case). The transmission curves on the other hand, show a high likelihood to be peaked around the average, as was expected from theoretical grounds[23]. The conductivity curves present a hybrid behavior between those two extremes.

To analyze these results, we proceeded in two different, but complementary ways. We directly fitted the curves to reasonable analytical forms, from which we could obtain statistical information. We also calculated the statistical moments $M_n$ from the "raw" data and then fitted $M_n$ to analytical expressions.

For the direct fit of the distributions we used $I(x) \propto e^{-x^{\mu}}$, $T(x) \propto e^{-(\frac{x-1}{\lambda})^2}$, $s(x) \propto e^{-x^{\gamma}}$, as suggested by the qualitative arguments presented above, where $\mu$, $\lambda$ and $\gamma$ are variational parameters that we adjusted to minimize the difference between the raw data and the analytical expression for all x (x=I/<I>, T/<T>, σ/<σ>, respectively). For the typical case of 60% coverage, we obtained $\mu$=1.0927, $\lambda$=0.208, $\gamma$=3.2192. We see that the Intensity follows almost a Rayleigh distribution (which corresponds to $\mu$=1), while T is peaked with a small dispersion, and σ shows a compressed exponential behavior in the sense that its tail has a much shorter range than that of a pure decaying exponential.

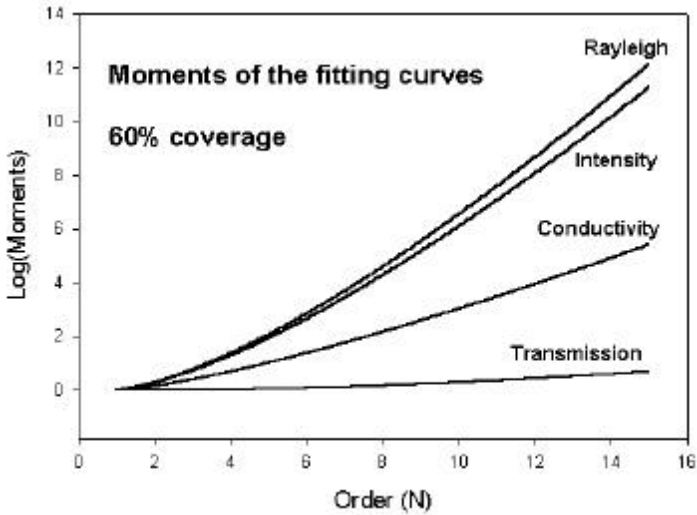
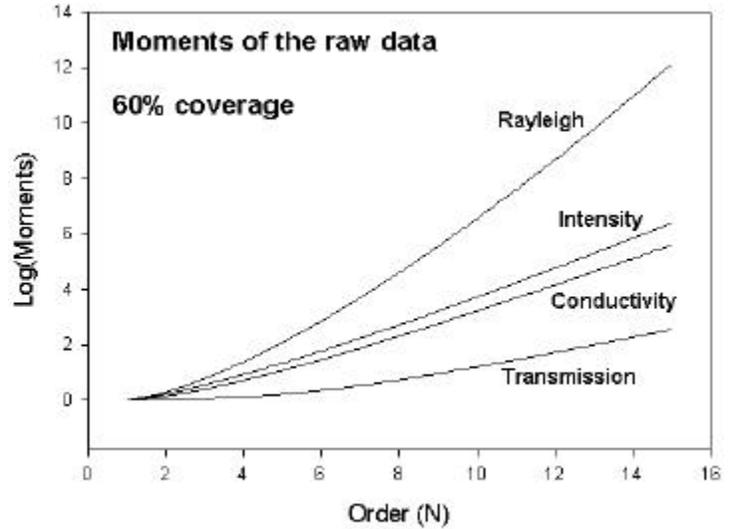

Figure 3. Moments of the adjusted distributions.      Figure 4. Moments of the raw data.

Fig. 3 shows that the moments of the distribution approach closely those of Rayleigh ($M_n$= n!), the moments of T are close to those of a δ-function ($M_n$ =1), a horizontal line in this plot), and the moments of σ are in an intermediate situation.

As mentioned above, we calculated independently the moments, $M_n$ from the raw data that gave rise to Fig 2. In Fig 4, we show plots for the first 15 moments of all relevant quantities for a system with a filling fraction of 60%.

For all values of the coverage considered in this work (20%-90%) the quantities <$I^n$>, <$T^n$>, and <$\sigma^n$> are always greater than unity (δ-function distribution) and less than n! (Rayleigh distribution). Since the δ-function distribution corresponds to no fluctuations we identify that with the limit of ballistic transport, while the Rayleigh distribution corresponds to an ideally diffusive regime; the results for the moments suggest that our system presents a mixed diffusive-ballistic regime.

To further support this hypothesis, we found that the above moment curves for I, T and σ for all filling fractions in the range 20-90%, can be well fitted by a single-parameter expression given by

$$M_n \equiv <X^n> = \sum_{m=0}^{n} \frac{(n!)^2 a^m (1-a)^{n-m}}{(m!)^2 (n-m)!}$$

where X = I, T or $\sigma$, and *a* is the fitting parameter.

Kogan and Kaveh[28], using diagrammatic arguments, argued that the above equation describes the moments $M_n = \langle I^n \rangle$, in the crossover between ballistic and diffusive transport where the parameter *a* controls the degree of diffusivity. Indeed, the two limiting cases a=0 and a=1 correspond, respectively, to the moments of the Rayleigh distribution and the $\delta$-function distribution.

As an example of the fitting, Fig. 5 shows the moments, $\langle T_n \rangle$ of the transmission for 60% coverage compared to the best fit curve, corresponding to a=0.965 which corresponds to a very peaked function, close to a $\delta$-function distribution.

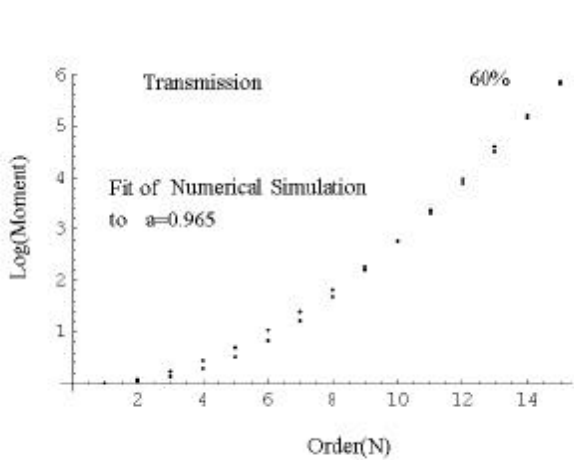

Figure 5.

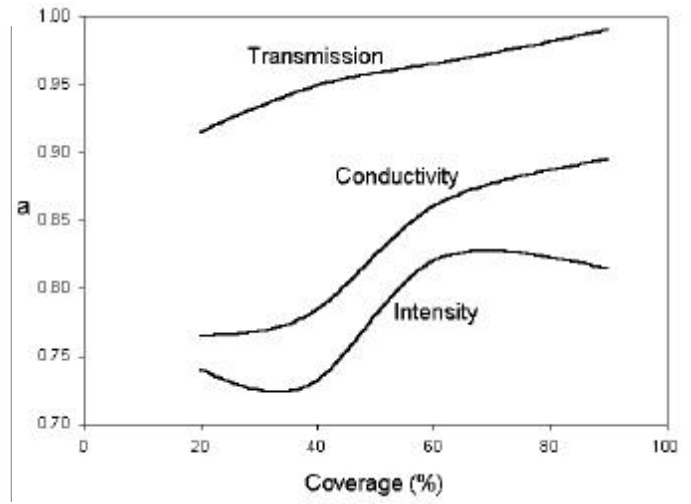

Figure 6.

The results of the fitting of the moments for increasing scatterer coverage are shown in Fig. 6. The Transmission becomes more $\delta$-like as the coverage increases. The Intensity remains in a clear hybrid diffusive-ballistic regime for all values of coverage although we could not identify a clear tendency in this range. In all cases the fitting parameter for the moments of the Conductivity remains at an intermediate value of those of the Intensity and Transmission.

## CONCLUSIONS

In this paper, we introduced a numerical model that allows for the calculation of the probability distributions of the intensity, transmission, and conductivity. Within this model, the moments of the Intensity fall between 1 and n!, suggesting that the transport regime is an admixture of ballistic and diffusive. This assumption is further supported by the fact that all the moments $M_n$ were appropriately fitted to independently derived analytical expressions valid for that type of crossover behavior. We expect that the transmission distribution should present a sharp peak both in the diffusive and ballistic regime. This is consistent with our results that show such a behavior for all coverages.

Our results also indicate that the intensity remains, for all values of the coverage we considered, in a crossover regime without a very clear tendency towards the diffusive limit.

**ACKNOWLEDGMENTS.**

This work is supported by Yeshiva University.

**REFERENCES**

[1] A.Z. Genack, Phys. Rev. Lett. **58**, 2043-2046 (1987)
[2] J. W. Strutt (Lord Rayleigh), Phil. Mag. **47**, 375 (1899)
[3] S. Chandrasekhar, "Radiative Transfer", Dover publishing, New York (1960)
[4] P. W. Anderson, Phys. Rev. **109**, 1492, (1958)
[5] E. Abrahams, P.W. Anderson, D. C. Licciardello, T.V. Ramakrishnan, Phys. Rev. Lett. **42**, 673 (1979)
[6] M. Stephen, Phys. Rev. Lett. **56**,1809-1810 (1986)
[7] M. P. van Albada, A. Lagendijk, Phys. Rev. Lett. **55**, 2692-2695 (1985)
[8] P.A. Lee, in STATPHYS 16, (H. E. Stanley, ed.), North Holland, Amsterdam, 169-174 (1986).
[9] G. Cwilich and Y. Fu, Phys. Rev. **B46**, 12015-12018 (1992)
[10] M. P. van Albada, B.A. van Tiggelen, Ad Lagendijk, Adriaan Tip, Phys. Rev. Lett. **66**, 3132-3135 (1991)
[11] A.D. Stone, Phys. Rev. Lett. **54**, 2692-2695 (1985)
[12] R.A. Webb, S. Washburn, C.P. Umbach, and R.B. Laibowitz, Phys. Rev. Lett. **54**, 2696-2699 (1985).
[13] N. Garcia, A. Z. Genack, Phys. Rev. Lett. **63**,1678-1681 (1989)
[14] C. Soukulis, in *Diffuse Waves in Complex Media* (J. P. Fouque ed.), Kluwer, 93-107 (1998).
[15] E. Yablonovitch, K.M. Leung, Nature **401**, 539 (1999)
[16] Mathias Fink, Physics Today **50**, 34 (1997)
[17] M. Campillo, L. Margerin and N.M. Shapiro, in *Diffuse Waves in Complex Media,* (J.P. Fouque ed.) Kluwer, 383-404 (1999)
[18] C.W.J. Beenakker, Phys. Rev. Lett. **81**, 1829-1832 (1998)
[19] D.S. Wiersma, M.P. van Albada, A. Lagendijk,, Nature **373**, 203-204 (1995).
[20] L. Wang, Appl. Optics **32**, 5043 (1993)
L. Wang, P.P. Ho, C. Liu, Science **253**, 769 (1991)
[21] R. Kaiser, in *Diffuse Waves in Complex Media*, (J.P. Fouque ed.), 249-288 , Kluwer (1999)
[22] D.S. Wiersma, P. Bartolini, Ad Lagendijk, R. Righini, Nature **390**, 671 – 673 (1997)
[23] E. Kogan, M. Kaveh, Phys. Rev. **B52**, R3813-R3815 (1995); E. Kogan, R. Baumgartner and R. Berkovits, Physica **A200**, 469 (1993); Kogan, M. Kaveh, R. Baumgartner, R. Berkovits, Phys. Rev. **B48**, 9404-9410 (1993)
[24] G. Cwilich, in OSA Proc. on Adv. in Opt. Imaging and Photon Migration (R. Alfano, ed.), **21** (1994)
[25] M. Stoytchev, A.Z. Genack, Phys. Rev. Lett. **79**, 309 (1997)
[26] P. Sebbah, O. Legrand, B. A. van Tiggelen, and A. Z. Genack, Phys. Rev. **E56**, 3619-3623 (1997)
[27] A.A. Chabanov, M. Stoychev and A.Z. Genack, Nature **404**, 850 – 853 (2000); M. Stoytchev, A.Z. Genack, Opt. Lett. **24**, 262 (1999)
[28] E. Kogan, M. Kaveh, Phys. Rev. **B22**, 16400 (1995)